\documentclass[letterpaper]{jpconf}
\usepackage{graphicx,amsmath}
\begin{document}
\title{The STAR W Program : New Results and Future Measurements}

\author{Joseph Seele for the STAR Collaboration}

\address{Massachusetts Institute of Technology}

\begin{abstract}
The production of $W^\pm$ bosons in longitudinally polarized p+p collisions at RHIC provides
a new means of studying the spin-flavour asymmetries of the proton sea quark spin distributions.
Details of the W$^\pm$ event selection through the $e^\pm$ decay channel at mid-rapidity are presented,
along with preliminary results for the production cross section and parity-violating single-spin
asymmetry, $A_L$, from the STAR Collaboration'Õs 2009 data set at $\sqrt{s} = 500$ GeV. Lastly, the expected
sensitivities for future running are discussed.

\end{abstract}

\section{Introduction}

High-energy polarized p+p collisions at $\sqrt{s} = 500$ GeV at RHIC provide a unique way
to study the partonic spin structure of the proton. Inclusive polarized deep-inelastic scattering (DIS)
experiments have shown that only $\sim$30\% of the spin of the proton is attributable to the polarization 
of the quarks \cite{Filippone:2001ux}. However, these inclusive measurements do not discern between  the various flavours of the 
quarks and their individual contributions. Semi-inclusive DIS measurements, however,
can achieve separation of the quark and anti-quark spin contributions by flavour \cite{Airapetian:2004zf}. The extracted anti-quark polarized Parton Distribution Functions (PDFs) have sizable
uncertainties compared to the well-constrained quark and anti-quark sums \cite{deFlorian:2009vb}.

$W^{-(+)}$ bosons are produced in p+p collisions, at leading order, through $\bar{u}+d (\bar{d}+u)$ interactions at the partonic
level and can be detected through their leptonic decays. The parity-violating nature of the weak
interaction gives rise to large single beam helicity asymmetries, $A_L$, which yield a
direct and independent probe of the quark and anti-quark polarized PDFs. The single helicity asymmetry is
defined as $A_L = \left(\sigma^+-\sigma^-\right)/\left(\sigma^++\sigma^-\right)$, where $\sigma^{+(-)}$ 
refers to the cross section when the helicity of
the polarized proton beam is positive (negative). Theoretical frameworks have been developed
to describe the production of W$^\pm$ bosons and their decay leptons in polarized p+p collisions \cite{deFlorian:2010aa,Nadolsky:2003ga}.

\section{Recent Measurements}

The measurements described in these proceedings were done using the STAR detector \cite{Ackermann:2002ad} at RHIC.
The used detector systems are the Time Projection Chamber
(TPC), which provides tracking of charged particles in a 0.5 T solenoidal magnetic field for
pseudorapidities of $|\eta|\leq1.3$ and the Barrel and Endcap Electromagnetic Calorimeters
(BEMC,EEMC), which are lead-scintillator sampling calorimeters covering $|\eta|\leq 1$ 
and $1.09\leq\eta\leq 2$, respectively. All three detectors provide 2$\pi$ coverage in azimuthal angle, $\phi$.
The data presented in this contribution, which correspond to 13.7 $\pm$ 0.3 (stat) $\pm$ 3.1(syst) pb$^{-1}$
of sampled luminosity, were accumulated in 2009 when a first
significant dataset was collected for polarized proton collisions at a centre of mass energy of $\sqrt{s} = 
500$ GeV. During this period, the beam polarizations averaged (38 $\pm$ 3)\% and (40 $\pm$ 5)\% for 
the two beams. 

In this analysis, only the $W^\pm\rightarrow e^\pm+\nu$ decay channel was considered. The electrons and positrons at mid-rapidity from these W$^\pm$ decays can be detected by a large transverse energy, $E_T$, peaked near $M_W/2$. In order to preferentially select the events containing a high energy $e^\pm$, a two-stage online BEMC trigger was required. First an event was required to satisfy a hardware
threshold corresponding to a transverse energy,  $E_T > 7.3$ GeV, in a single BEMC tower. Additionally, a software level trigger
then searched for a seed tower with $E_T > 5$ GeV and required that the maximum 2$\times$2 tower cluster
including that seed have an $E_T$ sum larger than 13 GeV. 

\begin{figure}[!t]
\begin{center}
\includegraphics[scale=0.70]{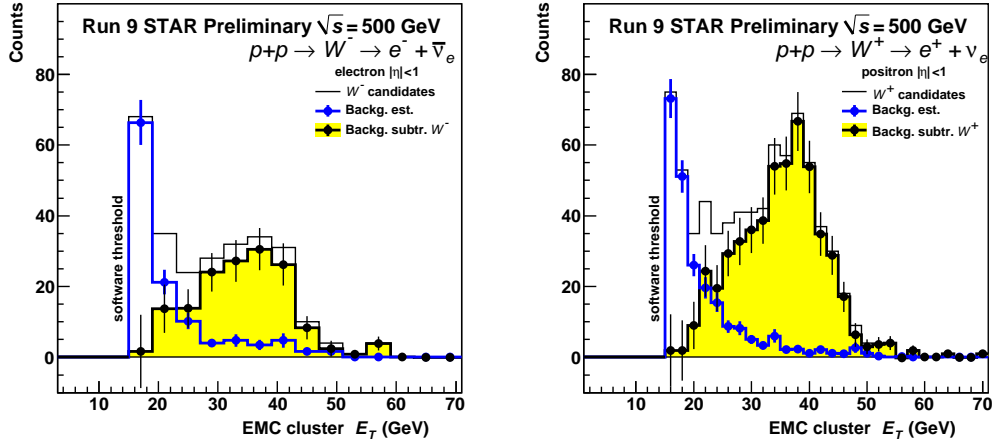}
\caption{The charge separated W$^\pm$ candidates distributions (black), the estimate of the background (blue) and the extracted signal (yellow).}
\label{fig-jacob}
\end{center}
\end{figure}

W$^\pm$ candidate events were selected offline based on kinematical and topological differences
between leptonic $W^\pm$ decay events and QCD background (e.g. di-jets) events. $W^\pm\rightarrow e^\pm+\nu$
decay events contain
a nearly isolated $e^\pm$ and an undetected neutrino opposite in azimuth, leading to a large missing
$E_T$. The selection proceeded in three stages : $e^\pm$ candidate identification, isolation, and background
rejection. An $e^\pm$ candidate is identified as any TPC track with
$p_T > 10$ GeV/c which originated from the event vertex with $|z| < 100$ cm, where $z$ is the 
direction along the beamline from the nominal interaction point. Furthermore, it was required that this track
points to a 2$\times$2 BEMC tower cluster with an $E_T$ sum, $E^{2\times 2}_T$, greater than 15 GeV and 
whose centroid was less than 7 cm from the extrapolated track.  Next, two isolation cuts are imposed on
the candidate. First, successful candidates were required to have an excess $E_T$ in the surrounding 4$\times$4 tower cluster be less than 5\% of $E^{2\times 2}_T$. Secondly, the candidates were required to have an excess EMC tower + TPC track $E_T$ sum of less than 12\% of $E^{2\times 2}_T$ within a cone radius $R = \sqrt{(\Delta\eta)^2+(\Delta\phi)^2}= 0.7$ of the candidate. Two background rejection cuts were applied. First, the magnitude of the vector $p_T$ sum
of the $e^\pm$ candidate $p_T$ vector and the $p_T$ vectors of all the reconstructed jets with thrust axes 
outside the $R = 0.7$ cone around the candidate was required to be greater than 15 GeV/c. The jets were 
reconstructed using the standard mid-point cone algorithm used in previous STAR jet 
measurements \cite{Abelev:2006uq,Abelev:2007vt}. Second, the away-side $E_T$,  which was defined to be  the EMC + TPC $E_T$ sum over the 
full range in pseudorapidity and having $\Delta\phi > 3\pi/4$ from the $e^\pm$ candidate track, was required to be less than 30 GeV.

\begin{figure}[!t]
\hfill
\begin{minipage}[!t]{.45\textwidth}
\begin{center}  
\includegraphics[scale=0.34]{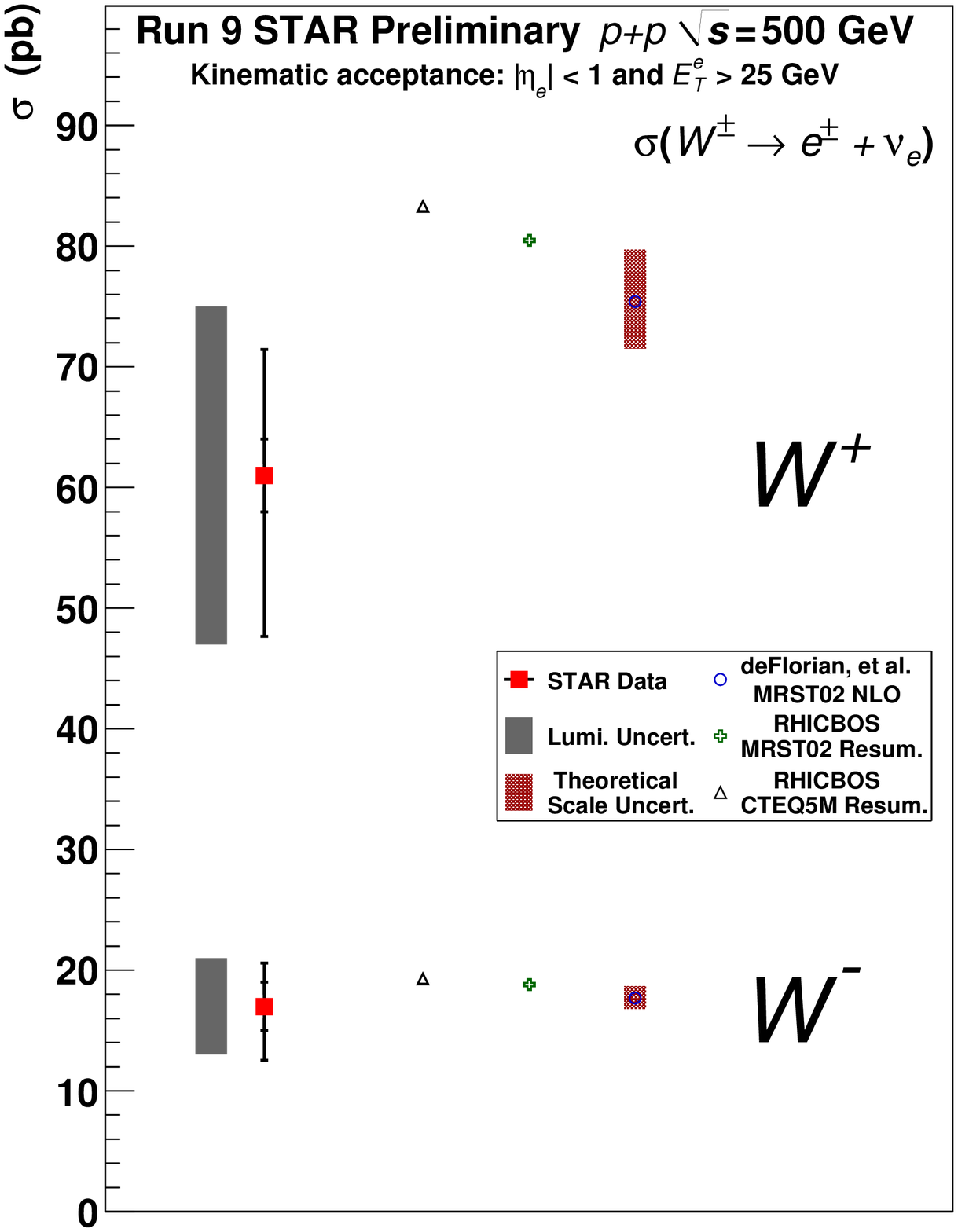}
\caption{The extracted cross sections for $W^\pm\rightarrow e^\pm+\nu$ in proton-proton collisions at $\sqrt{s}=500$ GeV. Also plotted are theoretical expectations \cite{deFlorian:2010aa,Nadolsky:2003ga} for a variety of PDFs.}
\label{fig-xsec}
\end{center}
\end{minipage}
\hfill
\begin{minipage}[!t]{.45\textwidth}
\begin{center}  
\includegraphics[scale=0.31]{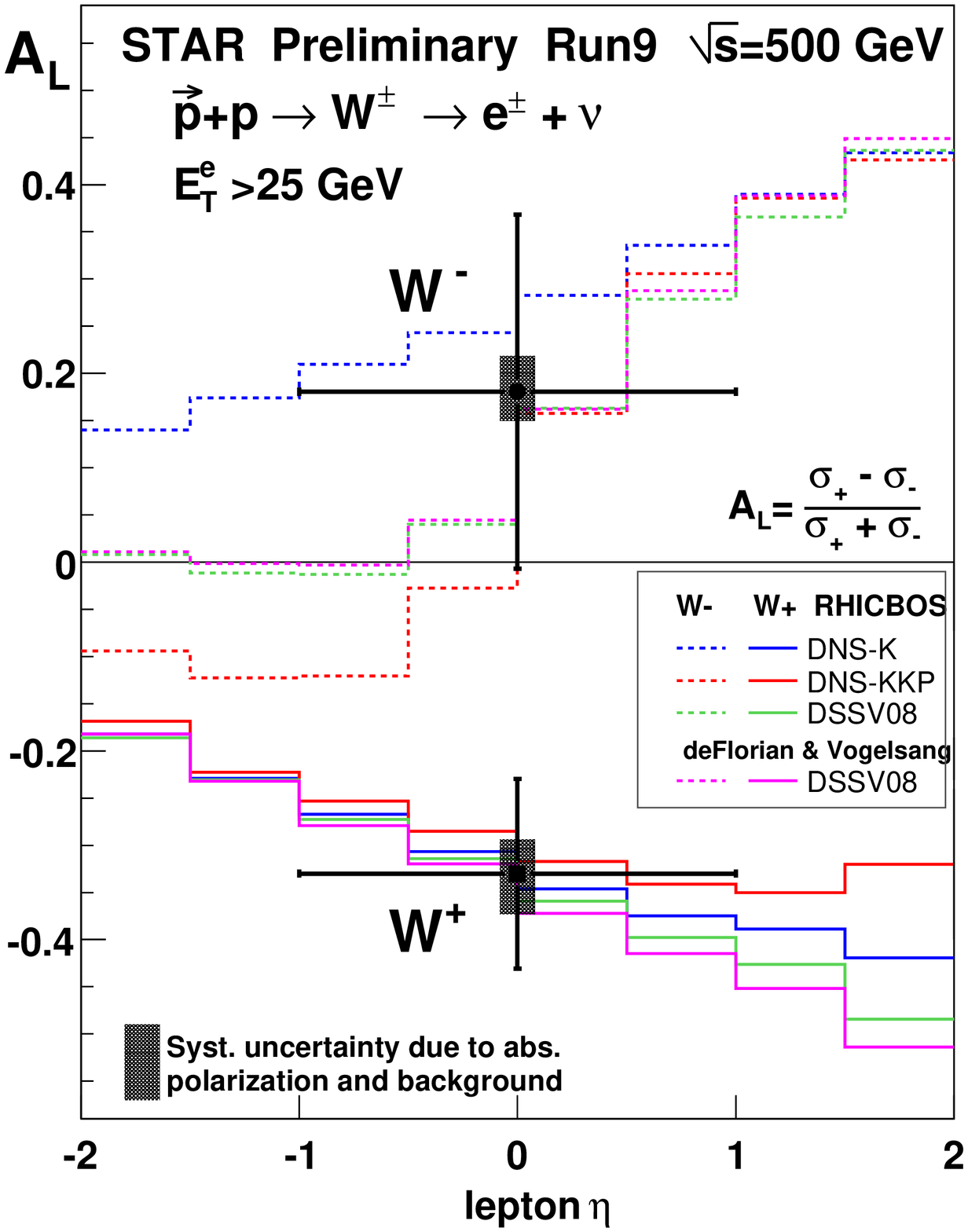}
\caption{The extracted single helicity asymmetries for $W^\pm\rightarrow e^\pm+\nu$ in polarized proton-proton collisions at $\sqrt{s}=500$ GeV. Also plotted are theoretical expectations \cite{deFlorian:2010aa,Nadolsky:2003ga} for a variety of polarized PDFs.}
\label{fig-al}
\end{center}
\end{minipage}
\hfill
\end{figure}

After these cuts were applied, the W$^\pm$ candidates (see figure \ref{fig-jacob}) show the characteristic Jacobian 
peak at $E^{2\times 2}_T\sim M_W$/2. The amount of background remaining in the W$^\pm$ candidate sample, after 
applying the selection criteria described above, was estimated from three contributions. The first contribution was
from $W^\pm\rightarrow\tau^\pm+\nu$ decay where the $\tau^\pm$ decays semi-leptonically to an $e^\pm$ and two neutrinos. 
This background was estimated using Monte Carlo simulation. 
Another contribution estimated the impact of the missing calorimetric coverage for $-2\leq\eta\leq -1.09$. To
determine this contribution to the background, the analysis was performed a second time with
the EEMC not used as an active detector. The difference in the W$^\pm$ candidate $E^{2\times 2}_T$
distribution with and without the EEMC included in the analysis was taken to be the estimate for the missing
calorimetric coverage. The third contribution was estimated by normalizing a data-driven background
shape to the remaining W$^\pm$ candidate signal, in the $E^{2\times 2}_T$ range below 19 GeV, after the first two 
background contributions were removed. This data-driven background  shape was obtained by inverting the last 
two requirements in the W$^\pm$ candidate selection, namely by requiring that the away-side $E_T$ be greater than 30 
GeV or the magnitude of the vector $p_T$ sum be less than 15 GeV/c. The total background was then subtracted 
from the W$^\pm$ candidate spectrum to yield the W$^\pm$ signal spectrum (see figure \ref{fig-jacob}). A systematic uncertainty 
for the background estimation was determined by varying the inverted cuts used to obtain the data-driven
background shape and by varying the range where the background shape was normalized to the
remaining W$^\pm$ candidate signal.

\begin{figure}[!t]
\begin{center}
\includegraphics[scale=0.32]{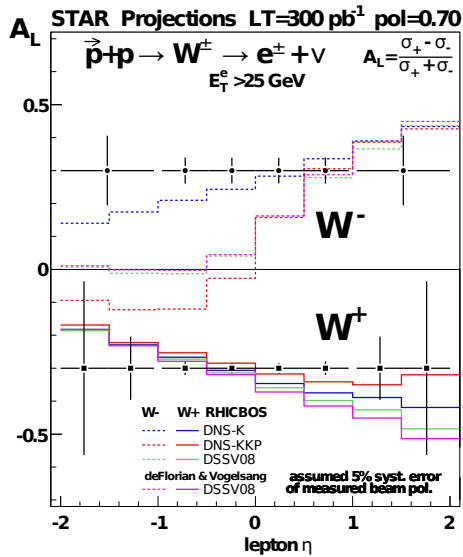}
\caption{Expected future sensitivities for the single helicity asymmetry for $W^\pm$ production in polarized proton-proton collisions at $\sqrt{s}=500$ GeV at RHIC. Also plotted are theoretical expectations \cite{deFlorian:2010aa,Nadolsky:2003ga} for a variety of polarized PDFs.}
\label{fig-al_proj}
\end{center}
\end{figure}

Preliminary results for the production cross sections and single helicity asymmetries of $W^\pm\rightarrow e^
\pm+\nu$ from events  with $|\eta|\leq 1$ and $E^{2\times 2}_T > 25$ GeV can be seen in figures 
\ref{fig-xsec} and \ref{fig-al}. The measured values are $\sigma(W^+\rightarrow e^++\nu)$ = 61 $\pm$ 3 (stat.) 
$^{+10}_{-13}$ (syst.) $\pm$14 (lumi.) pb and $\sigma(W^-\rightarrow e^-+\bar{\nu})$ = 17 $\pm$  2
(stat.) $^{+3}_{-4}$ (syst.) $\pm$ 4 (lumi.) pb. The measured asymmetries are $A^{W^+}_L = -0.33 \pm 
0.10$ (stat.) $\pm$ 0.04 (syst.) and $A^{W^-}_L = 0.18 \pm 0.19$ (stat.) $^{+0.04}_{-0.03}$ (syst.). These 
results are consistent with the theoretical calculations using the same acceptance and polarized PDFs which have been constrained by polarized DIS experiments.

\section{Future Measurements}

In the future, STAR plans to take more data at $\sqrt{s}=500$ GeV. This data will place a clean constraint on the quark and anti-quark polarized PDFs of the proton. The expected sensitivities for the single helicity asymmetry as a function of pseudorapidity can be seen in figure \ref{fig-al_proj}. The calculation in the figure assumed our demonstrated efficiency and background rejection in the mid-rapidity region while in the forward and backward rapidity regions, a worse signal to background ratio was assumed. The uncertainties are for 300 pb$^{-1}$ of sampled luminosity and 70\% average polarization which is the the goal of the longitudinal $\sqrt{s}=500$ GeV spin program at RHIC.

\section{Summary}

The STAR Collaboration has taken its first data with longitudinally polarized proton beams at $\sqrt{s}=500$ GeV in 2009. First results on the mid-rapidity cross section and single helicity asymmetry for $W^\pm$ were presented.  This opens a new era in the study of the spin-flavour structure of the proton 
based on the production of $W^{\pm}$ bosons. 

\section*{References}
\bibliography{iopart-num}

\end{document}